\documentstyle[12pt]{article}
 \newcommand{ \beq}{ \begin{equation}}
 \newcommand{ \enq}{ \end{equation}}
 \newcommand{ \lan}{ \langle}
 \newcommand{ \ran}{ \rangle}
\def\cu#1{``#1''}
 
\begin{document}
\begin{flushright}
                hep-th/9411059 \\
\end{flushright}
\title{Non--commutative Group Manifolds}

\author{Z. Hasiewicz and P. Siemion\thanks{
Supported by KBN grant No 2P 302 087 06 }
\\
\\
Institute of Theoretical Physics\\
 University of Wroc{\l}aw, pl. M. Borna 9 \\
 50--205 Wroc{\l}aw, Poland
 }
\date{}
\maketitle

 \begin{abstract}
We show that a chiral sector of a symplectic group manifold
possesses a symmetry similar to, but somewhat weaker than the
Lie--Poisson one.
 \end{abstract}

 \section{Splitting of a cotangent bundle}
 \medskip \par
For the dynamical systems with a high degree of symmetry it is natural to
try to parametrize as much of the phase space, as possible by the global
constants of motion. \par
In the case of geodesic motion on a semisimple group manifold (with the
Hamiltonian being given by the quadratic Casimir invariant) all global
constants of motion are described by the momentum mappings corresponding
to the natural left and right actions of the group on its cotangent
bundle. The equivariance of the momentum mappings allows one to split
the phase space $T^*G$ into the sectors corresponding to the types of
coadjoint orbits. In the case of compact groups this decomposition is
quite simple as one has the unique type of the orbits of maximal dimension.

It has been shown  \cite{my} that in such a case one can represent $T^*G$
by the Cartesian square of $G  \times { \cal W}$ divided by suitable relations.
(${ \cal W}$ stands for some chosen Weyl chamber, used to parametrize the
space of coadjoint orbits).  \par
We call $G  \times { \cal W}$ a Chiral sector.
 The canonical symplectic form of the cotangent bundle pulls back
 onto the product of the two sectors as a difference $  \Omega_L -  \Omega_R$
 of two components, each component living on one sector. \par
Each component is an exact two--form, giving each sector a
structure of a symplectic manifold.  \par

Thus the classical model can now be quantized in two different ways:
 \begin{enumerate}
 \item One can quantize  the cotangent bundle. This is straightforward
and yields known results. The operators corresponding to the matrix elements
of any representation of $G$ commute.

 \item One can quantize each of the sectors separately. It is much more
complicated, but results in a very interesting class of non--commutative
($ \cal C^*$ for compact groups) algebras describing a new class of quantum
group
manifolds. In each sector there is a non--commutative spectrum generating
algebra (SGA). \\
 \end{enumerate}
The following diagram summarizes these ideas:

 
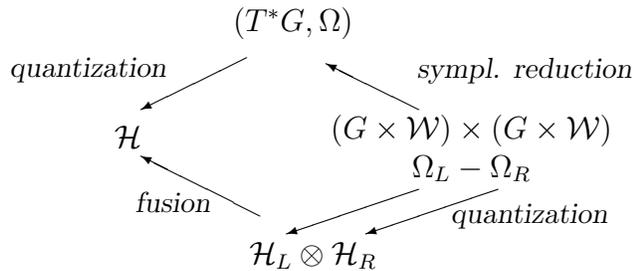
\begin{picture}(130,95)(0,10)
 \put(100,90){ \makebox(35,10)[b]{$      (T^*G, \Omega)   $}}
 \put(35,45){ \makebox(40,10){$  { \cal H} $}}
 \put(105,0){ \makebox(40,10){$   { \cal H}_L  \otimes { \cal H}_R  $}}
 \put(165,40){ \makebox(40,10){$  \matrix{
        (G  \times { \cal W})  \times (G  \times { \cal W})  \cr
         \Omega_L -  \Omega_R } $}}
 \put(20,70){ \makebox(40,10){ \small  \sl quantization }}
 \put(185,70){ \makebox(40,10){ \small  \sl sympl. reduction }}
 \put(60,20){ \makebox(20,10){ \small  \sl fusion}}
 \put(185,15){ \makebox(40,10){ \small  \sl quantization}}

 \put(100,80){ \vector(-2,-1){40}}
 \put(165,60){ \vector(-2,1){35}}
 \put(105,20){ \vector(-2,1){45}}
 \put(165,30){ \vector(-3,-1){50}}
 \put(195,30){ \vector(-3,-1){50}}
 \end{picture}
 \\[0.5cm]

The detailed description of the above procedure called
chiral splitting and fusion can be found in  \cite{my}. \\

 \section{Symplectic structure of a sector}
The symplectic structure of the chiral sector can be most transparently
described in terms of Chevalley basis of ${ \cal G}$.

  In case of compact groups the space of coadjoint orbits can be conveniently
parametrized by choosing some Weyl chamber $W$ in ${ \cal G}$ (i.e. the dual
of a Cartan subalgebra divided by the Weyl group of its discreet
symmetries). $W$ intersects each regular orbit exactly once.

We shall use the following notation \cite{Hum}. The set of simple roots
dual to the chosen Weyl chamber is $ \Delta$, the
set of roots of the Lie algebra is $ \Phi$, and the set of positive roots
 is
$ \Phi_+$. The element of the Cartan subalgebra $K$--dual to the root
$ \beta$
is  $t_{ \beta}=i [e_ \beta,e_{- \beta}]$. In addition we introduce
$ \theta^{ \alpha_i}$, the  one--form dual to the simple root
$t_{ \alpha_i}$
and $ \omega^\beta$, the  left invariant one--form dual to the root
vector
$e_ \beta$. Finally, $w_i$ is the  coordinate in the Weyl chamber in the
basis dual to the one formed by $t_{ \alpha_i}$.
 \par
 The left component
(the symplectic form of the left sector) is given by
 \beq
 \Omega_L= \sum_{ \alpha_i  \in  \Delta} dw_i  \wedge \theta^{ \alpha_i}
+i  \sum_{ \beta \in \Phi_+}  \lan w,t_ \beta  \ran  \omega^\beta \wedge
 \omega^{- \beta}.
 \label{sectsym}
 \enq
It has a global symplectic potential:
 \beq
  \Omega_L = d  \sum_i w_i  \theta^{ \alpha_i}.
 \enq
For the 'right' component the expressions are analogous.
 \par
The symplectic structure gives Poisson brackets of matrix elements
in arbitrary representations of $G$ as:
  \beq
  \{T_1  \otimes T_2 \}_M(g) = (T_1  \otimes T_2)(g) r_{12}(w)
  \enq
 where
  \beq
  r_{12}(w) =   \sum_{ \beta \in \Phi_+} \frac{i}{ \lan w,t_ \beta  \ran}
    [ \tau_1(e_{- \beta}) \otimes  \tau_2(e_ \beta)-
       \tau_1(e_{ \beta}) \otimes  \tau_2(e_{- \beta})].
 \label{r}
 \enq
together with
 \beq
 \{w_i,T \}(g)=T(g)  \tau(t_{ \alpha_i}). \label{PB wi M}
 \enq
($T$ and $\tau$ label representations  of $G$, and their differentials,
respectively.)
 \section{Example: G=SU(2)}
In the fundamental representation $T_f$:
 \beq
  T_f(g) =  \pmatrix{ a & -b^*  \cr b & a^* }   \; ;  \; w  \in { \bf R}_+
 \enq
 \beq
         a^{*}a+b^{*}b = 1 .
 \enq
 \begin{eqnarray}
 \{ a^{*}, a  \}_M = {i \over w} b b^{*} &,& \quad  \{ \, a , b  \, \}_M = 0
\qquad ,
 \nonumber \\
 \{ a , b^{*} \}_M = {i \over w} a b^{*} &,& \quad
   \{ b , b^{*} \}_M = - {i \over w}a a^{*} ,
 \label{Poisson a b}  \\
 \{ a , w \,  \}_M = - i a  \ &,& \quad  \{ a^{*}, w  \}_M = i a^{*} ,
 \nonumber \\
 \{ b , w  \, \}_M = - i b  \ &,& \quad  \{ b^{*}, w  \}_M = i b^{*} ,
 \label{Poisson w}
 \end{eqnarray}
Geometric Quantization \cite{Wood} \cite{Snia} of this structure gives:
 \beq
         { \hat a}^{ \dagger}{ \hat a}+{ \hat b}^{ \dagger}{ \hat b} = 1 .
 \enq
 \beq
     \Delta = 1 - { \hat a}{ \hat a}^\dagger - { \hat b}{ \hat b}^\dagger  \;(=
 \hbar  \hat w ^{-1} )
 \enq
 \beq
  \label{niekom}
  \matrix{
 (1+ \Delta) { \hat a}^\dagger{ \hat a} = { \hat a}{ \hat a}^\dagger +  \Delta
\cr
  \cr
 (1+ \Delta) { \hat b}^\dagger{ \hat b} = { \hat b}{ \hat b}^\dagger +  \Delta
\cr
  \cr
 (1+ \Delta) { \hat a}^\dagger{ \hat b} = { \hat b}{ \hat a}^\dagger   \cr
  \cr
 ab = ba }
 \enq
We can say that the above relations define the structure of non--commutative
group manifold, namely \cu{quantum} $S^3$.
In the representation Hilbert space there is a unique 'vacuum' state
$ \varphi_o$ satisfying
 \beq
           a^\dagger  \varphi_o = 0 = b^\dagger  \varphi_o
 \enq
We can expect that the procedure of geometric quantization will enable us
to describe the corresponding non--commutative manifolds for all compact
groups.

 \section{Trace of the Right Action Symmetry}
 \par
By performing the chiral splitting we have broken the right symmetry of
the dynamical system. This happened because we have chosen some fixed Weyl
chamber in order to parametrize the coadjoint orbits. \\
The symmetry is restored after fusion of both sectors. One may ask however
whether there is a trace of the broken symmetry in the chiral sector? \\
The natural right action of $G$ on the sector is given by:
 \beq
  (G \times W)  \times G  \ni (g,w,h)  \buildrel{R} \over \mapsto (gh^{-1},w)
   \in G  \times W.
 \enq
We assume that the acting group $G$ is equipped with the bracket $ \{.,. \}_G$,
such that the above action preserves the quadratic Poisson brackets { \it
for the group elements} in the chiral sector. We should stress that we do
not demand the preservation of the brackets of functions on $ \cal W$
with the group elements. \par
The equation for the bracket $ \{.,. \}_G$ reads:
 \beq
   \label{eqq}
 \matrix{  \{T_1  \otimes T_2 \}_M(gh^{-1}) =  \cr  \cr
 =   \{T_1  \otimes T_2 \}_M(g) (T_1  \otimes T_2)(h^{-1}) +
    (T_1  \otimes T_2)(g) \{T_1  \otimes T_2 \}_G(h^{-1}) }
  \enq
In terms of the Poisson tensor ${cal P}_G$ corresponding to $\{.,.\}_G$,
the unique solution to (\ref{eqq}) is:
 \beq
   { \cal P}_G = L^*_g r - R_g^{*-1} r,
 \enq
Where $r$ is that of (\ref{r}). \par
Question: is ${ \cal P}_G$ Lie--Poisson? It has  the familiar form of
a 'Sklyanin bracket', but does the bivector $r$ satisfy YBE ? \\

For $SU(2)$ the answer is affirmative.
In this case we obtain a family of Lie Poisson structures labeled by the
Weyl chamber parameter $w$ :
 \begin{eqnarray}
 \{ \,  \alpha  ,  \beta  \,   \}_G &=&- {i \over w}  \alpha  \beta  \, \quad ;
\quad
 \{  \alpha ^{*},  \beta ^{*} \}_G = {i \over w}  \alpha ^{*} \beta ^{*} ;
 \nonumber \\
 \{  \alpha  ,  \beta ^{*} \}_G &=& - {i \over w}  \alpha  \beta ^{*}  \quad ;
 \quad
 \{  \alpha  ,  \alpha ^{*} \}_G = {2i \over w}  \beta ^{*} \beta  ;
 \nonumber \\
 \{  \beta ^{*},  \beta   \}_G  &=& 0 \quad .\phantom{{2i \over w}  \beta ^{*}
\beta}
 \label{Woron}
 \end{eqnarray}

For the compact groups of higher rank the tensor ${ \cal P}_G$ is {\it not}
a Poisson one as it breaks the Jacobi identity for the corresponding bracket
of functions.
It is not clear to us at this moment how to realize the quantum version
of the above symmetry as we don`t know which of the relations
(\ref{niekom}) are independent. We hope to get back to this problem
in a forthcoming paper. \\

 \end{document}